\documentclass[prd,english,preprintnumbers,amsmath,amssymb,nofootinbib,twocolumn,superscriptaddress,aps,10pt]{revtex4-1}

\pdfoutput=1

\usepackage[utf8]{inputenc}
\usepackage{graphicx}
\usepackage{bbm}
\usepackage{amssymb}
\usepackage{amsmath}
\usepackage{tabularx}
\usepackage{slashed}

\usepackage{array}
\newcommand{\PreserveBackslash}[1]{\let\temp=\\#1\let\\=\temp}
\newcolumntype{C}[1]{>{\PreserveBackslash\centering}p{#1}}
\newcolumntype{R}[1]{>{\PreserveBackslash\raggedleft}p{#1}}
\newcolumntype{L}[1]{>{\PreserveBackslash\raggedright}p{#1}}

\usepackage{multirow}
\usepackage{colortbl}
\definecolor{kugray5}{RGB}{224,224,224}

\usepackage{xspace}
\usepackage[export]{adjustbox}  

\usepackage{dsfont}
\usepackage{babel}

\usepackage{hyperref}

\def\0#1#2{\frac{#1}{#2}}

\def\s0#1#2{\mbox{\small{$ \frac{#1}{#2} $}}}

\newcommand{\be}{\begin{eqnarray}}
\newcommand{\ee}{\end{eqnarray}}

\newcommand{\nn}{\nonumber }

\newcommand{\beq}{\begin{equation}}
\newcommand{\eeq}{\end{equation}}
\newcommand{\bea}{\begin{eqnarray}}
\newcommand{\eea}{\end{eqnarray}}

\def\eq#1{\eqref{#1}}
\def\eqref#1{(\ref{#1})}

\makeatother
\begin{document}

\title{Chiral Susceptibility in (2+1)-flavour QCD}

\author{Jens Braun}
\affiliation{Institut f\"ur Kernphysik (Theoriezentrum), Technische Universit\"at Darmstadt, 
D-64289 Darmstadt, Germany}
\affiliation{ExtreMe Matter Institute EMMI, GSI, Planckstra{\ss}e 1, D-64291 Darmstadt, Germany}
\author{Wei-jie Fu} 
\affiliation{School of Physics, Dalian University of Technology, Dalian, 116024, P.R. China}
\author{Jan M. Pawlowski}
\affiliation{Institut f\"ur Theoretische Physik, Universit\"at Heidelberg, Philosophenweg 16, 69120 Heidelberg, Germany}
\affiliation{ExtreMe Matter Institute EMMI, GSI, Planckstra{\ss}e 1, D-64291 Darmstadt, Germany}
\author{Fabian Rennecke}
\affiliation{Physics Department, Brookhaven National Laboratory, Upton, NY 11973, USA}
\author{Daniel Rosenbl\"uh}
\affiliation{Institut f\"ur Kernphysik (Theoriezentrum), Technische Universit\"at Darmstadt, 
D-64289 Darmstadt, Germany}
\author{Shi Yin}
\affiliation{School of Physics, Dalian University of Technology, Dalian, 116024, P.R. China}

\begin{abstract}
We calculate chiral susceptibilities in (2+1)-flavour QCD for different masses of the light quarks 
using the functional renormalisation group (fRG) approach to first-principles QCD. We follow the evolution of the chiral 
susceptibilities with decreasing masses as obtained from both the light-quark and the reduced quark condensate. 
The latter compares very well with recent results from the HotQCD collaboration for pion masses $m_{\pi}\gtrsim 100\,\text{MeV}$.
For smaller pion masses, the fRG and lattice results are still consistent. 
In particular, the estimates for the chiral critical temperature are in very good agreement. We close by discussing different extrapolations to the chiral limit. 
\end{abstract}

\maketitle

%
\section{Introduction}\label{sec:intro}
The phase structure of QCD probed with heavy-ion collisions  
is well described by (2+1)-flavour QCD. While the charm, bottom and top quarks are too heavy to significantly add to the dynamics of the system, the 
dynamics of the strange and, most importantly, of the light up and down quarks determine the rich phase structure in particular at large densities. Recently, 
functional methods for first-principles QCD have made significant progress in the description of 
this regime, see Refs.~\cite{Fu:2019hdw, Braun:2019aow} for functional renormalisation group studies (fRG) and, e.g., Refs.~\cite{Fischer:2018sdj, Isserstedt:2019pgx, Gao:2020qsj} for Dyson-Schwinger studies. 
Still, in the high-density regime the systematic error of the current computations grows large. This asks for both, 
systematically improved computations and a better error control. In turn, lattice simulations are obstructed by the sign problem at finite density and either 
rely on Taylor expansions at vanishing chemical potential~\cite{Allton:2002zi,Gavai:2003mf,Allton:2003vx,Kaczmarek:2011zz,Bazavov:2018mes} 
or on analytic continuations from imaginary to real quark chemical potential~\cite{deForcrand:2002hgr,DElia:2002tig,DElia:2007bkz,deForcrand:2010he,Laermann:2013lma,Bonati:2014kpa,Philipsen:2016hkv}. 
In summary, this suggests a two-tier strategy to tackle the high-density regime by direct systematically improved functional computations and a quantitative access to the zero-density limit. 

Interestingly, the mass dependence of the phase structure at vanishing density 
can potentially constrain the phase structure at large density.
For instance, low energy effective theory computations indicate that the chiral phase transition temperature in the limit of 
massless up and down quarks is a possible upper bound for the transition temperature at the critical endpoint, 
see Ref.~\cite{Ding:2020rtq} for a recent review.
Accordingly, the nature of the chiral transition in QCD with three quark flavours is very actively researched. 
For sufficiently small masses of the three quarks, one expects a finite mass range with a first-order chiral transition~\cite{Pisarski:1983ms}. 
Interestingly, this first-order regime may even extend to the limit of infinitely heavy strange quarks, see, e.g., Refs.~\cite{Philipsen:2016hkv,Cuteri:2017gci,Cuteri:2017zcb}. 
This intricate question regarding the existence and range of such a regime  
is tightly connected to the fate of the axial~$U_{\rm A}(1)$ anomaly at finite temperature: depending on the strength of the associated $U_{\rm A}(1)$ breaking, the 
phase transition may indeed be of first order~\cite{Pisarski:1983ms, Rennecke:2016tkm, Pisarski:2019upw}.

For physical masses of the three quarks, the 
transition in (2+1)-flavour QCD from a low-temperature hadronic phase to a high-temperature 
quark-gluon plasma phase has been found in lattice and functional QCD studies to be a crossover, 
see, e.g., Refs.~\cite{Aoki:2006we,Aoki:2006br, Bonati:2018nut, Borsanyi:2018grb, Bazavov:2018mes, Guenther:2018flo, Ding:2019prx} 
for lattice studies and Refs.~\cite{Fu:2019hdw, Braun:2019aow, Fischer:2018sdj, Isserstedt:2019pgx, Gao:2020qsj} 
for functional studies. 

In the chiral limit of the light quarks, the 
critical behaviour is controlled by the three-dimensional ($3d$) $O(4)$ universality class, if the anomalous breaking of the $U_{\rm A}(1)$ symmetry 
is sufficiently strong. In turn, if the $U_{\rm A}(1)$ symmetry is effectively restored sufficiently close 
to the chiral transition, the critical behaviour may no longer be 
controlled by the $3d$ $O(4)$ universality class~\cite{THOOFT1986357,Grahl:2013pba,Pelissetto:2013hqa,Sato:2014axa}. 
Within a very recent lattice QCD study investigating pion masses in the range of $50\,\text{MeV}\lesssim m_{\pi} \lesssim 160\,\text{MeV}$ with 
a physical strange quark mass,
the scaling properties of the chiral susceptibility are now found to be compatible with the $3d$ $O(4)$ universality class~\cite{Ding:2019prx}. 
An extrapolation to the chiral limit of the light quarks leads to $T_{\text{c}}=132^{+3}_{-6}\,\text{MeV}$ for the chiral critical temperature~\cite{Ding:2019prx}. 

The reconstruction of the chiral critical temperature from an extrapolation to the chiral limit is in general a non-trivial task 
as it is affected by non-universal aspects, such as the order of the transition and the 
dependence of the pseudocritical temperature on the pion mass. Moreover, 
the definition of a pseudocritical temperature is not unique. Indeed, the strength of the pion-mass 
dependence of the pseudocritical temperature is different for different definitions.   
Assuming that the chiral phase transition is of second order in the limit of massless up and down quarks, 
it follows from universal scaling arguments ~\cite{Goldenfeld:1992qy} that 
the pion-mass scaling of the pseudocritical temperature defined as the position of the peak of the chiral susceptibility 
is controlled by the critical exponents of the underlying universality class.
Unfortunately, the size of the scaling regime is also a non-universal quantity. 
These statements also hold for other definitions of the pseudocritical temperature and 
it is also reasonable to expect that the size of the scaling regime is of the same order for 
different definitions, provided that they
rely on properties of the chiral susceptibility. 

In low-energy effective theories of QCD it has been found that the pseudocritical temperature 
defined as the position of the peak of the chiral susceptibility scales roughly linearly 
over a wide range of pion masses which appears 
compatible with scaling arguments at first glance~\cite{Berges:1997eu,Braun:2005fj,Braun:2010vd}. 
Even more, it was found that the results for the suitably rescaled chiral order parameter 
fall almost on one line~\cite{Braun:2010vd} for~$m_{\pi}\gtrsim 75\,\text{MeV}$, seemingly suggesting scaling behavior. However, a comparison of these 
results with the corresponding scaling function extracted within the model studies  
exhibits 
clear deviations from scaling. A detailed analysis then revealed that the 
size of the actual scaling regime is very small, i.e., scaling behavior of the chiral susceptibility and the chiral order parameter 
is only observed for very small pion masses,~$m_{\pi}\lesssim 1\,\text{MeV}$, see Ref.~\cite{Braun:2010vd}.

Our present first-principles fRG study corroborates these findings in low-energy 
effective theories: the actual scaling regime in QCD is indeed small, with a conservatively estimated 
upper bound of~$m_{\pi}\approx 30\,\text{MeV}$. In addition, our present work shows
that the glue dynamics softens the strong dependence 
of the pseudocritical temperature 
on the pion mass observed in low-energy effective theories of QCD. 
In fact, it is found to be in very good agreement with recent lattice QCD results~\cite{Ding:2019prx}. 
Moreover, we shall discuss different extrapolations to the chiral limit, leading us consistently to~$T_{\text{c}}\approx 142\,\text{MeV}$ 
for the critical temperature, see also Figs.~\ref{fig:renorm_cs} and \ref{fig:tpcmpi} below.

This work is organized as follows: In Sect.~\ref{sec:fw}, we briefly discuss the methodological framework of our present 
study. Our results for the chiral susceptibility 
as obtained from the light-quark condensate are  presented in Sect.~\ref{sec:res}. There, we also
show a comparison of these results with those for the susceptibility extracted from the reduced condensate, also used in lattice computations. The results 
for the reduced condensate are then compared to lattice QCD data~\cite{Ding:2019prx}, including a discussion of the 
dependence of the pseudocritical temperature on the pion mass. Our conclusions can be found in Sect.~\ref{sec:conc}.

\section{Condensates \& functional QCD}\label{sec:fw}
In this section we discuss different chiral condensates and the 
associated susceptibilities which we compute to access the 
mass dependence and scaling of the pseudocritical temperature. 
We also briefly introduce 
the fRG approach to QCD used for this computation.
The basis for our present study is discussed in detail in~Ref.~\cite{Fu:2019hdw}.

\subsection{Chiral condensates}\label{sec:condintro}
In order to obtain the (chiral) susceptibility in $(2+1)$-flavour QCD for various (current) quark masses~$m_{q_i}^{0}$, we have to compute the chiral condensates~$\Delta_{q_i}$ associated with the 
three quark flavours~$q_i$. Here, $q_i=u,d,s$ refers 
to the up, down, and strange quark, respectively. The~$\Delta_{q_i}$ can be obtained from the logarithmic derivative of the 
thermodynamic grand potential~$\Omega$ with respect to the corresponding current quark mass, 
\begin{align}
\Delta_{q_i} =&\, m_{q_i}^{0}\frac{\partial \Omega(m_q;T,\mu_{u},\mu_{d},\mu_{s})}{\partial m_{q_i}^{0}}\nn\\[1ex]
=&\, m_{q_i}^{0}\frac{T}{V} \int_0^{\frac{1}{T}}{\rm d}\tau\int_{V}{\rm d}^{3}x\left\langle \bar{q}_i(\tau,\vec{x}^{\,})q_i(\tau,\vec{x}^{\,})\right\rangle\,.
\label{eq:Condensate}
\end{align}
Here,~$T$ is the temperature and~$V$ is the spatial volume.
The logarithmic derivative with respect to the current quark mass is taken since~$\Delta_{q_i}$ then carries 
the same scaling properties as the grand potential~$\Omega$. Consequently, it is not sensitive to details of the setup 
that typically change the precise value of the current quark mass, in particular the renormalisation scheme, 
see Ref.~\cite{Fu:2019hdw} for a discussion. 
In the present work, we only consider the zero-density limit and therefore 
we set the quark chemical potentials to zero,~$\mu_u=\mu_d=\mu_s=0$. Thus, the quark condensates are only functions of the temperature 
and the current quark masses from here on. 
Moreover, we use identical current masses for the two light 
quarks, $m^0_u=m^0_d=m^0_l$. This allows to define the light quark condensate~\mbox{$\Delta_l=\Delta_u=\Delta_d$}. 

The computation of the quark condensates via the expectation value in the last line 
of Eq.~\eqref{eq:Condensate} requires renormalisation and hence the result depends on the renormalisation procedure. 
Both, the necessity for renormalising the operator, and the renormalisation scheme dependence is removed when considering finite difference of 
chiral condensates. Two possible choices are the \textit{renormalised} and the \textit{reduced} condensate.  
Both are  commonly used in lattice QCD studies and have also been studied in functional approaches to QCD. 
The fRG approach naturally provides a renormalised finite expression for the condensate~$\Delta_{q_i}$ as it is based 
on a finite free energy and hence the current mass derivative in Eq.~\eq{eq:Condensate} is finite, for more details 
see Sect.~\ref{sec:fRG} and Ref.~\cite{Fu:2019hdw}. For our study of the chiral phase transition, the condensate~$\Delta_{l}$ is therefore 
the key observable in the present work since it has the smallest systematic error
within the truncation used for the computation.

 The \textit{renormalised} condensate associated with the light-quark flavours can be 
 defined as 
\begin{align}
\Delta_{l,R}(T)= \frac{1}{{\mathcal N}_{R}} \left(\Delta_{l}(T) - \Delta_{l}(0) \right)\,. \label{eq:DeltaRen}
\end{align}
The normalisation constant ${\mathcal N}_R$ is at our disposal: in the following, it is chosen 
independent of the current masses and is typically used to 
render $\Delta_{l,R}(T)$ dimensionless. 
Note that, by including a $m_l^0$-dependence of the form~${\mathcal N}_R\sim m_l^0/m_s^0$, we recover the definition of the renormalised 
condensate conventionally employed in lattice QCD studies, see, e.g., Ref.~\cite{Bazavov:2011nk}. 

The \textit{reduced} condensate~$\Delta_{l,s}$ is a combination of the light-quark condensate~$\Delta_l$ and the 
strange quark condensate~$\Delta_{s}$,  
\begin{align}
\Delta_{l,s}=\frac{1}{{\mathcal N}_{l,s}}\left(  \Delta_{l}(T) - \left(\frac{m_l^{0}}{m_s^{0}}\right)^2 \Delta_{s}(T) \right)\,,
 \label{eq:DeltaRed}
\end{align}
where the definition of the $m_l^{0}$-independent normalisation 
constant~${\mathcal N}_{l,s}$ is again irrelevant for our discussion of the susceptibilities below. 
Instead, if we choose~${\mathcal N}_{l,s}$ to be $m_l^0$-dependent,~${\mathcal N}_{l,s}=( \Delta_{l}(0) - (m_l^0/m_s^0)^2 \Delta_{s}(0) )^2$, 
we arrive at the standard lattice definition of the reduced condensate, see, e.g., Ref.~\cite{Bazavov:2011nk}. 
Alternatively, we could simply choose~${\mathcal N}_{l,s}\sim m_l^0/m_s^0$ which 
yields the observable defined in Ref.~\cite{Ding:2019prx} (up to numerical factors) to compute susceptibilities. 

The subtraction in Eq.~\eqref{eq:DeltaRed} renders the reduced condensate finite as in the case for the 
renormalised condensate. However, the systematic error of results of such a 
light-strange quark mixture is a combination of that in the strange and in the light quark sector and 
requires a quantitative treatment of both. Consequently, it is affected by larger systematic errors in our present fRG study than  
the light-quark condensate~$\Delta_{l}$. 

The corresponding susceptibilities are readily obtained from all three condensates. We define them as follows,  
\begin{align}
\chi_{M}^{(i)}(T) = -\frac{\partial}{\partial m_l^{0}}\left(\frac{\Delta_{i}(T)}{m_l^0}\right)\,, 
\label{eq:MagSus}
\end{align}
where $(i)=(l),\,(l,R),\,(l,s)$. We shall refer to these susceptibilities as 
light-quark susceptibility, renormalised susceptibility, and reduced susceptibility, respectively. 
Leaving an overall normalisation aside, 
it follows from the definition of the light-quark condensate~$\Delta_{l}$ and 
the renormalised condensate~$\Delta_{l,R}$ that the associated susceptibilities only 
differ by a temperature-independent shift. From our discussion above, it moreover 
follows that our definition of the reduced susceptibility matches the one used in lattice 
studies~\cite{Ding:2019prx}.

By multiplying 
the magnetic susceptibilities~\eq{eq:MagSus} with $(m_l^0)^2$, 
they also carry the scaling properties of the grand potential and there is no 
dependence on the renormalisation procedure left. This is in one-to-one 
correspondence to the lack of renormalisation scheme dependence of $\Delta_l$ 
defined by the logarithmic $m_l^0$-derivative of the grand potential, and to that 
of the
renormalised and reduced condensates.
However, for the sake of a straightforward comparison with the lattice results from Ref.~\cite{Ding:2019prx}, 
we have not included these factors in Eq.~\eq{eq:MagSus}. Moreover, for a comparison of the susceptibilities 
for different pion masses (or current quark masses) as well as for a comparison with results from other methods, 
it is convenient to 
normalise the magnetic susceptibilities $\chi_{M}^{(i)}(T)$ with the respective peak value for the physical pion mass $m_{\pi}\approx 140\,\text{MeV}$,
\begin{align}
\bar\chi_{M}^{(i)}=\max_T \chi_{M}^{(i)}(T)\Big|_{m_{\pi}=140\,\text{MeV}}\,,
\label{eq:SusNorm}\end{align}
where again $(i)=(l),\,(l,R),\,(l,s)$. Thus, we have $\chi_{M}^{(i)}(T)/ \bar\chi_{M}^{(i)}=1$
at the peak position in case of the physical pion mass.  
The size and evolution of the increasing peak towards the 
chiral limit gives a rough estimate for the ``distance" to criticality.

\subsection{Functional renormalisation group approach}\label{sec:fRG}
In this work, we use the fRG approach for the computation of 
the light-quark and reduced susceptibilities from first principles. In this approach, 
quark, gluon and hadron correlation functions of QCD are computed from functional relations that are 
derived from the flow equation for the \textit{finite} effective action $\Gamma$. 
Accordingly, the \textit{finite} effective action is easily accessible in this approach, and is self-consistent.
In particular, it automatically encodes the same RG scheme as the correlation functions.   

The thermodynamic grand potential $\Omega$ is then given by the effective action evaluated on the 
equations of motion (EoM), i.e., the ground state: $\Omega=(T/V)\Gamma|_\textrm{EoM}$. Hence, the fRG approach 
provides a finite thermodynamic grand potential~$\Omega$. This leads to a finite quark condensate $\Delta_{q_i}$ within the RG scheme used for the computation of the correlation functions, see Eq.~\eqref{eq:Condensate}. 

The computation of the susceptibilities in our fRG study requires the computation of the light-quark and reduced quark condensates
for various temperatures and quark masses.
To this end, we have to follow the RG flow from the classical QCD action in the ultraviolet to the long-range (infrared) limit where the dynamics 
is effectively described by hadronic degrees of freedom rather than quarks and gluons. To facilitate the description of the transition between the degrees of freedom in the ultraviolet and infrared limit, we employ \textit{dynamical hadronisation}  
techniques~\cite{Gies:2001nw, Gies:2002hq, Pawlowski:2005xe, Floerchinger:2009uf, Fu:2019hdw}, see  
Refs.~\cite{Gies:2002hq, Braun:2008pi, Mitter:2014wpa, Braun:2014ata, Rennecke:2015eba, Cyrol:2017ewj, Fu:2019hdw} for their application to QCD. 
The chiral susceptibilities can then be obtained with two different -- formally equivalent procedures -- from Eq.~\eq{eq:Condensate}, 
both of which are detailed below around Eq.~\eq{eq:MagSusAn} as their comparison provides an important self-consistency and reliability check for our present truncation. 

Our present study has been done within the {\it fQCD collaboration}~\cite{fQCD}, and is a follow-up of a recent work~\cite{Fu:2019hdw} 
within this collaboration. 
It also builds on previous advances made within this collaboration, see, e.g.,
Refs.~\cite{Pawlowski:2014zaa, Mitter:2014wpa, Braun:2014ata, Fu:2015naa, Rennecke:2015eba, Cyrol:2017ewj, Fu:2018swz, Braun:2019aow}. Therefore, we refrain from showing the 
flow equations required to compute the light-quark and reduced quark condensate
because of the size of this set 
of equations. All these equations are derived, documented and discussed in detail in Ref.~\cite{Fu:2019hdw}.
\begin{figure}[t]
	\centering
	\includegraphics[width=1\linewidth]{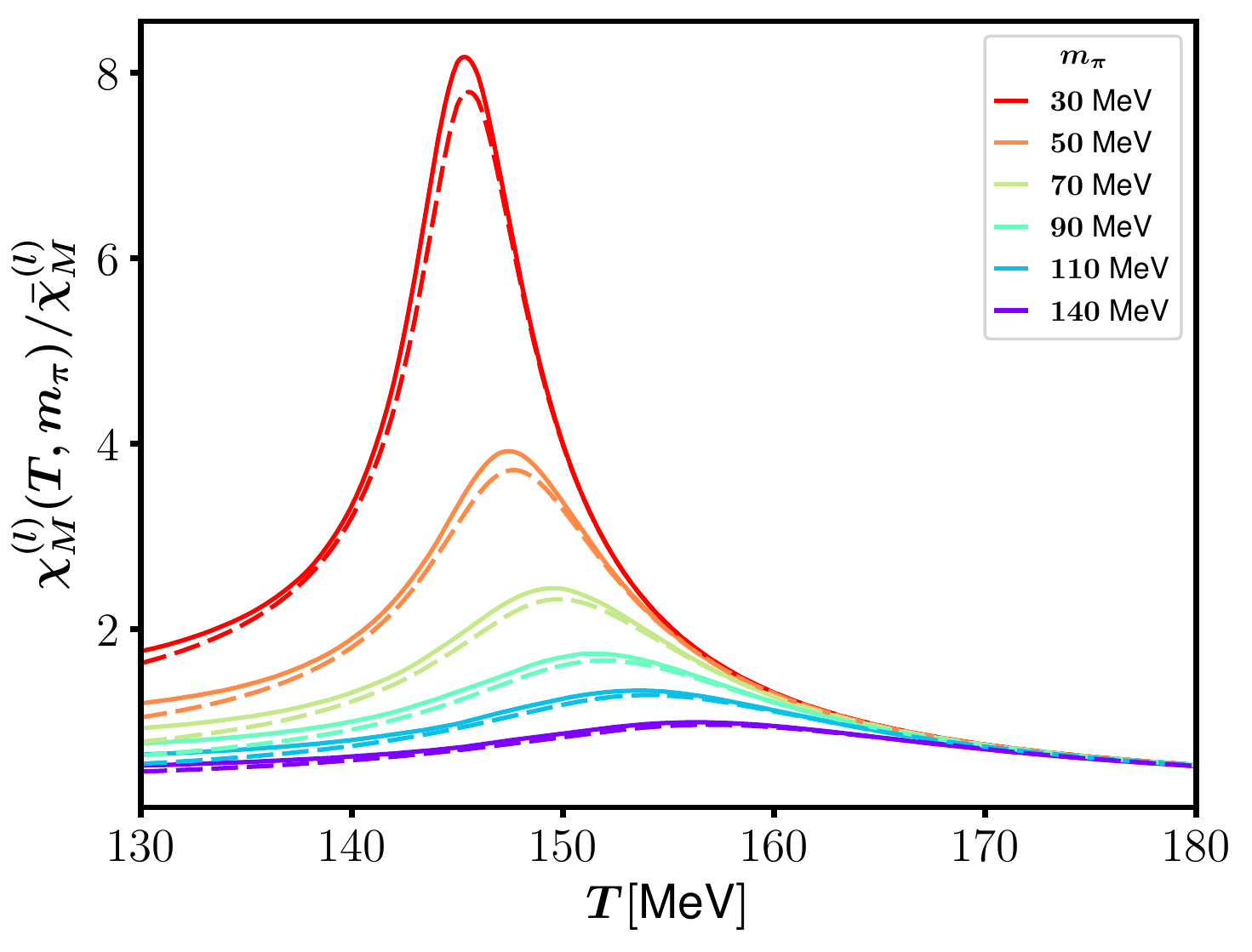} 
	\caption{Light-quark susceptibility for different pion masses 
	as obtained from two independent methods within our present truncation. The dashed 
	lines correspond to the results obtained via the relation~\eqref{eq:MagSusAn}. The solid lines are associated 
	with the results computed by taking a derivative of our numerical data set for the light-quark condensate with respect to $m^0_l$.
	The normalisation is given by the maximum of the susceptibility at the physical pion 
	mass, see Eq.~\eq{eq:SusNorm}. We observe that the results from the two methods agree very well for~$m_{\pi}\gtrsim 30\,\text{MeV}$. 
	This is particularly true for the respective peak positions, see main text for details.
	\label{fig:renorm_cs_ee}}
\end{figure}

We only would like to discuss two aspects of  
our computation which are particularly relevant for the estimate of the systematic error of our results:\\[-2ex]
  
First, we only take into account the sigma--pion channel in the computation 
of the order-parameter potential, and do not allow for an effective $U_{\rm A}(1)$ restoration at, e.g., high 
temperatures. Thus, we tacitly assume that the chiral transition falls into the $3d$ $O(4)$ universality class. 
This assumption is based on the fact that Fierz-complete finite-temperature studies 
of the chiral transition indeed indicate that the sigma--pion interaction channel is by far most 
dominant close to the chiral phase transition at vanishing baryon density~\cite{Braun:2019aow}. 
A discussion of whether this dominance is strong enough relative to $U_{\rm A}(1)$-restoring 
channels for keeping QCD in the $3d$ $O(4)$ universality class is beyond the scope of the present work.
However, a recent lattice study suggests that this may indeed be the case and that the transition is 
of second order in the chiral limit~\cite{Ding:2019prx}. \\[-2ex]

Second, in order to compute the chiral order-parameter potential, we employ a 
Taylor expansion about the RG scale-dependent minimum of the effective action. 
An inclusion of a finite pion mass into the flow equations 
then tends to stabilise this expansion~\cite{Braun:2004yk,Braun:2007td}. However, 
it becomes numerically unstable 
for (very) small pion masses. 

In the present work, we employ an error estimate to evaluate the 
self-consistency and reliability of our results in case of small pion masses. To this end, we 
compute the light-quark susceptibility with two different methods. 
Within the first method, we directly perform the derivative of the condensate $\Delta_{l}$ with respect to the 
light-quark current mass:
\begin{align}\label{eq:MagSusAn}
\!\!\chi_{M}^{(l)}(T) \!=\! \left(\frac{T}{V}\right)^2\!\!\int_{x,y}\!\!\langle \bar{q}_l(x)q_l(x)\, \bar{q}_l(y)q_l(y)\rangle \!-\! \left(\frac{\Delta_{l}}{m_l^0}\right)^2.
\end{align}
This expression can be computed directly from the grand potential~$\Omega$ for a \textit{given} current quark mass~$m_l^0$. Indeed, 
it is directly related to the screening mass of the $\sigma$-meson, see Refs.~\cite{Braun:2007td,Fu:2019hdw} for details.

Within the second method, the light-quark susceptibility~$\chi_{M}^{(l)}$ is simply 
obtained by computing the light-quark condensate as a 
function of the temperature and the light quark mass $m_l^0$, and then taking a numerical derivative of this data set 
with respect to $m^0_l$. Note that we also compute the reduced condensate in this way. 
\begin{figure*}[t]
	\centering
	\includegraphics[width=0.48\linewidth]{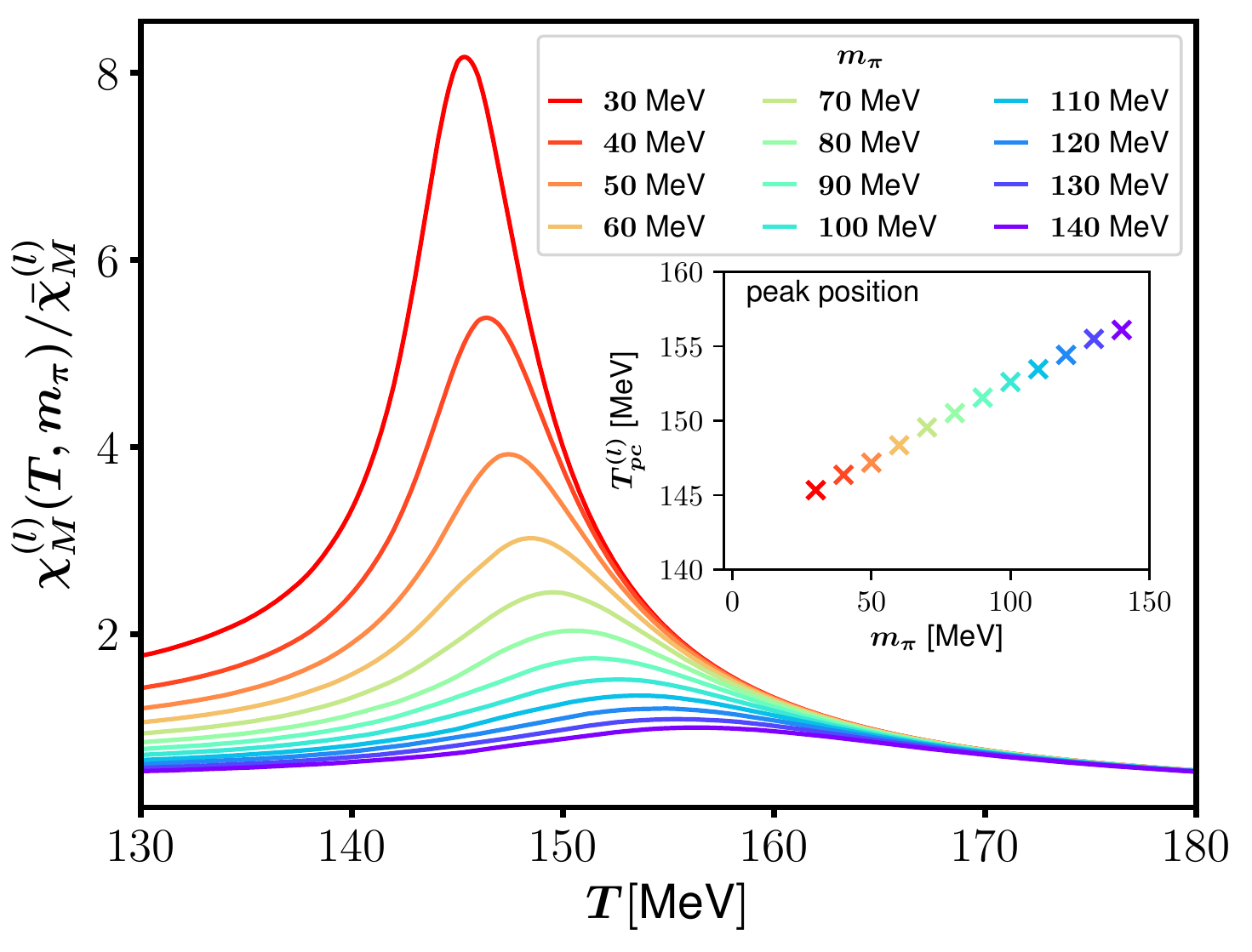} 
	\includegraphics[width=0.475\linewidth]{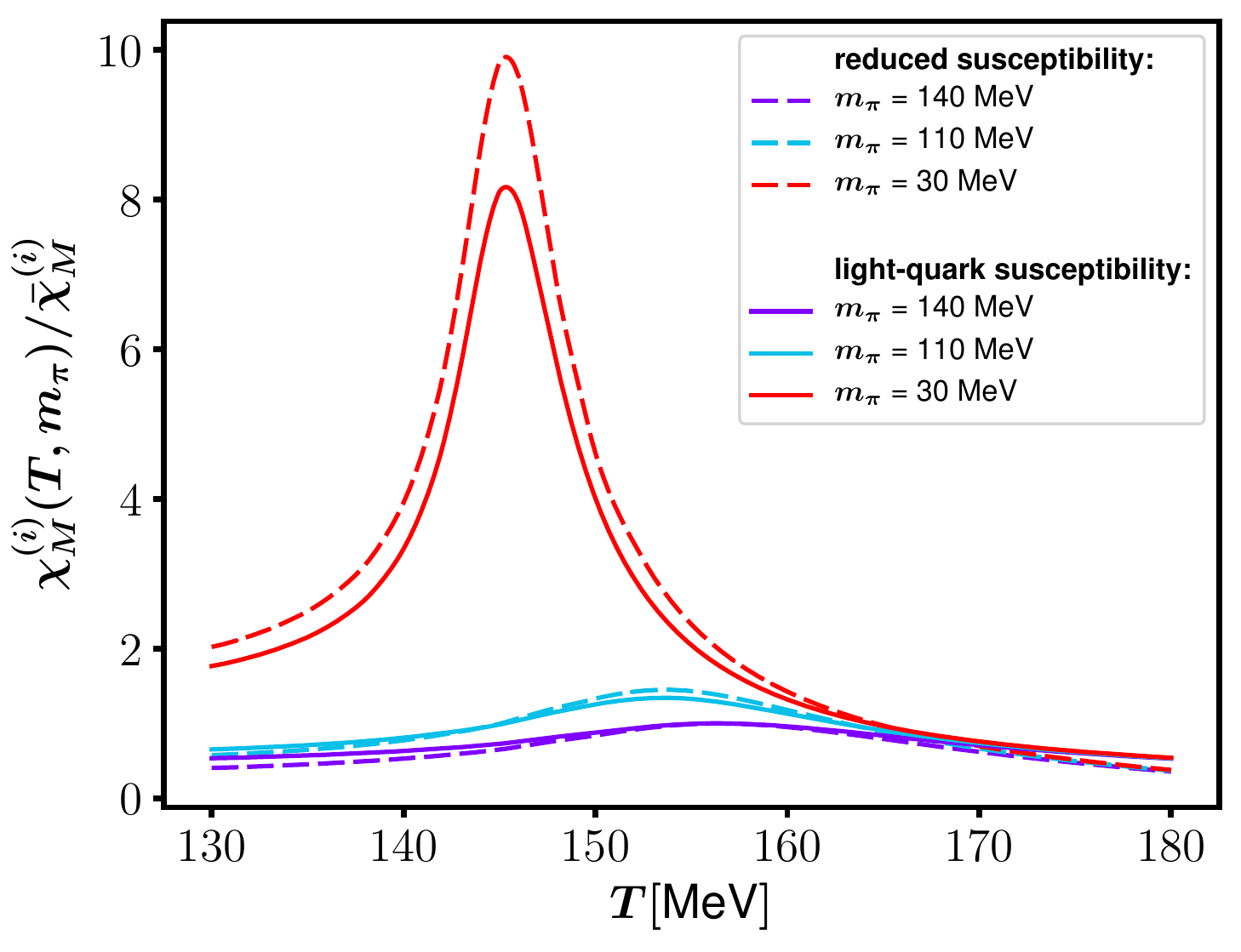} 
	\caption{Left panel: Light-quark susceptibility~$\chi_M^{(l)}$ as a function 
		of the temperature. The inset shows the peak positions of the depicted susceptibilities as a function of the pion mass. 
		Right panel: Comparison of the light-quark susceptibility~$\chi_M^{(l)}$ and the reduced susceptibility~$\chi_M^{(l,s)}$ 
		as a function 
		of the temperature. The normalisations 
		are the maxima of the respective susceptibilities at the physical pion mass, see Eq.~\eq{eq:SusNorm}.} 
	\label{fig:renorm_cs}
\end{figure*}

Evidently, both methods should give the same results for the light-quark susceptibility, provided that no approximations are involved. 
Therefore, deviations 
give access to the reliability of the underlying approximations. A comparison of results from the two methods is 
depicted in Fig.~\ref{fig:renorm_cs_ee}. From this, we conclude that our present approximation is 
trustworthy for pion masses $m_{\pi}\gtrsim 30\,\text{MeV}$.
In particular, the peak positions obtained from the two methods 
are in very good agreement with each other 
in this pion-mass regime. In fact, they only differ by about~$1\,\text{MeV}$. Based on this comparison, we consider it sufficient 
to show in the following only those results for the light-quark susceptibility 
which have been obtained by taking a derivative of our numerical data set for the light-quark condensate with respect to $m^0_l$.
As the light-quark condensate is part of the 
reduced condensate, we may assume that the uncertainty for the 
pseudocritical temperatures extracted from the reduced susceptibilities is the same as in the case of the light-quark susceptibilities. 
However, the systematic error  
is likely to be bigger in this case, as already indicated in Sect.~\ref{sec:condintro}.
Regarding smaller pion masses, we add that an extension in this direction (including the chiral 
 limit) is possible but requires either an expansion of the effective action about a scale-independent point~\cite{Pawlowski:2014zaa} 
 or the use of, e.g., recently developed techniques to access the full order-parameter potential~\cite{Grossi:2019urj}.

\section{Results}\label{sec:res}
Let us now discuss the susceptibilities associated with the light-quark and reduced condensate 
for various temperatures 
and pion masses. To this end, we shall keep the strange quark 
mass fixed at its physical value and only vary the light quark mass.
As already discussed in the previous section, the renormalised condensate only differs 
from the light-quark condensate by a temperature-independent shift. This implies that 
the peak positions of these two susceptibilities are the same for a given pion mass. Therefore, we shall not 
further discuss the renormalised susceptibility in this section. 

Susceptibilities as introduced in the previous section are of great interest as 
their maxima can be used to define pseudocritical temperatures. 
From general scaling arguments~\cite{Goldenfeld:1992qy}, it 
then follows that the  
pseudo\-critical temperatures extracted from the light-quark and reduced susceptibility scale as 
\begin{align}
T_{\text{pc}}^{(i)}(m_{\pi})\approx T_{\text{c}} + c_{(i)}\,m_{\pi}^{p}\,,
\label{eq:TpcRfct}
\end{align}
at least within the scaling regime. Here,~$(i)=(l),\,(l,s)$. The chiral critical temperature is 
given by~$T_{\text{c}}$ in Eq.~\eqref{eq:TpcRfct}. 
The quantity $c_{(i)}$ is a   
non-universal 
constant depending on the susceptibility under consideration whereas the exponent~$p$ can be related to the universal critical exponents~$\beta$ 
and~$\delta$, $p= 2/(\beta\delta)$. 
The relation \eq{eq:TpcRfct} follows from the fact 
that the position of the peak of the susceptibility as a function of the scaling variable~$z=t/h^{1/(\beta\delta)}$ is constant in the scaling regime. 
Here,~$t = (T-T_{\rm c})/T_0$ is the reduced temperature with~$T_0$ being a suitably chosen 
normalisation factor and~$h=H/H_0$ is the symmetry breaking field normalised by a suitably chosen 
normalisation~$H_0$. In the present case, $H$ can be identified with the current mass of the light quarks~$m_l^{0}$ 
which, in turn, is directly related to the pion mass via~$m_{\pi}^2\sim m_l^{0}$. 

For example, employing the critical exponents of the $3d$ $O(4)$ universality class~\cite{Guida:1998bx,Kanaya:1994qe,Hasenbusch:2000ph}, 
we have~$p\approx 1.08$. Based on previous fRG studies of critical exponents~\cite{Tetradis:1993ts,Bonanno:2000yp,Litim:2010tt,Balog:2019rrg}, however, 
we expect~$p$ to be slightly smaller in our present study. 
In any case, this suggests an almost linear dependence 
of the peak positions of the susceptibilities on the pion mass, at least within the scaling regime.
Note that the size of the latter is not universal but depends on the details of the 
theory under consideration.

In Fig.~\ref{fig:renorm_cs} (left panel), we show our results for the light-quark susceptibilities~$\chi_M^{(l)}(T)$ 
as a function of the temperature~$T$, where we have normalised the 
susceptibilities with~$\bar \chi_M^{(l)}$, i.e.,~the value of~$\chi_M^{(l)}(T)$ for~$m_{\pi}=140\,\text{MeV}$
evaluated at its maximum, see Eq.~\eq{eq:SusNorm}.
As expected, we find that the susceptibility increases when the pion mass is decreased. Indeed, by decreasing 
the pion mass, we approach the chiral limit associated with a diverging susceptibility at the chiral phase transition temperature~$T_{\rm c}$. 

Our results for the pseudocritical temperature indeed appear to depend
almost linearly on the pion mass, see Fig.~\ref{fig:renorm_cs} (left panel). 
Fitting the scaling relation~\eqref{eq:TpcRfct}
to our numerical results for~$T_{\text{pc}}^{(l)}(m_{\pi})$ for~$m_{\pi}=30, 35, 40, \dots, 140\,\text{MeV}$, 
we obtain~$T_{\text{c}}\approx 141.4_{-0.5}^{+0.5}\,\text{MeV}$, $c_{(l)}\approx 0.19_{-0.05}^{+0.05} \,\text{MeV}^{1-p}$, 
and~$p\approx 0.88_{-0.05}^{+0.05}$. At this point, we would like remind the reader that the renormalised 
susceptibility obeys the same temperature dependence as the light-quark susceptibility. Therefore, the 
pseudocritical temperatures extracted from these two susceptibilities are identical.  
\begin{figure*}[t] 
\centering
\includegraphics[width=0.50\linewidth]{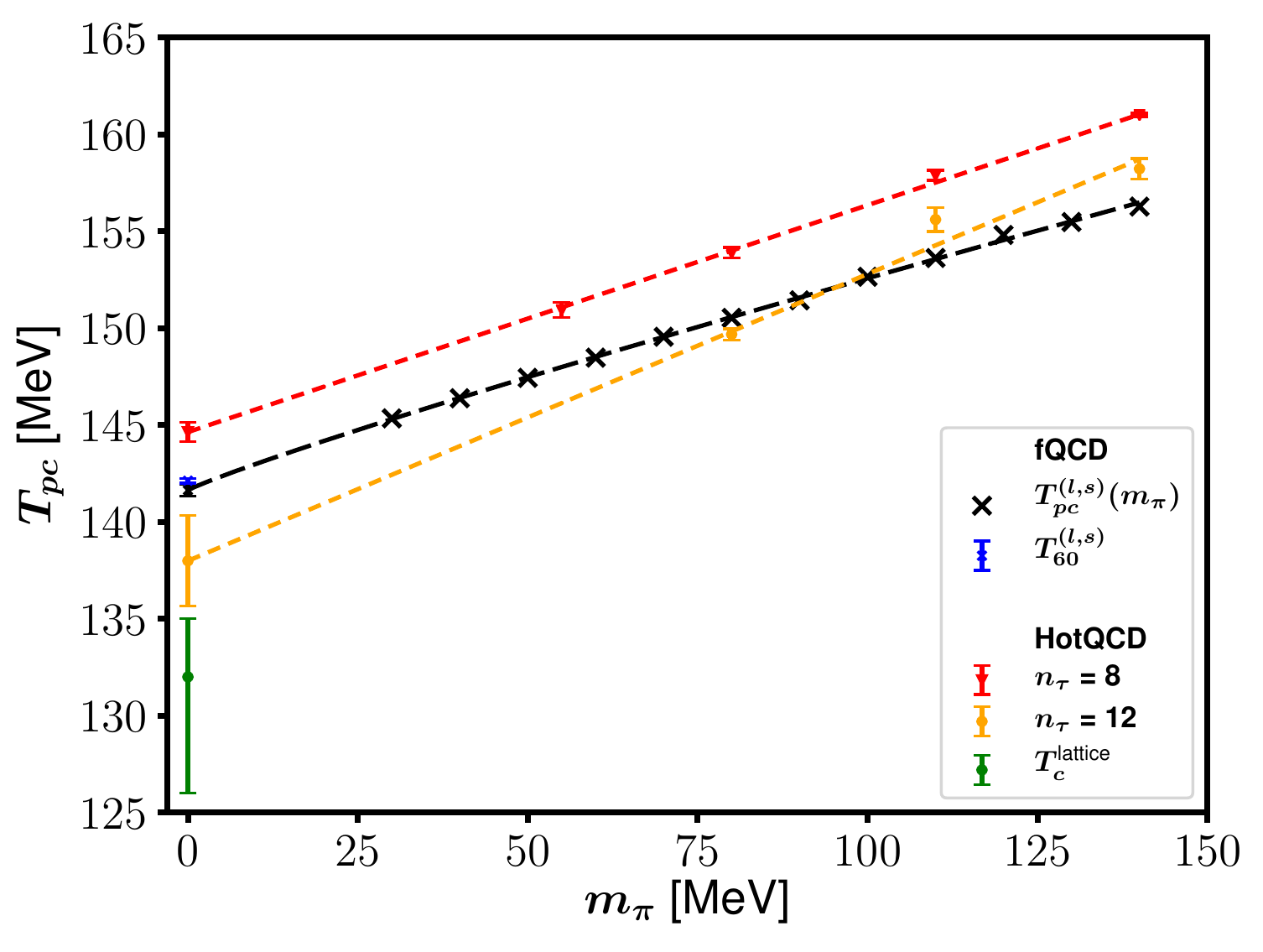}  
\includegraphics[width=0.475\linewidth]{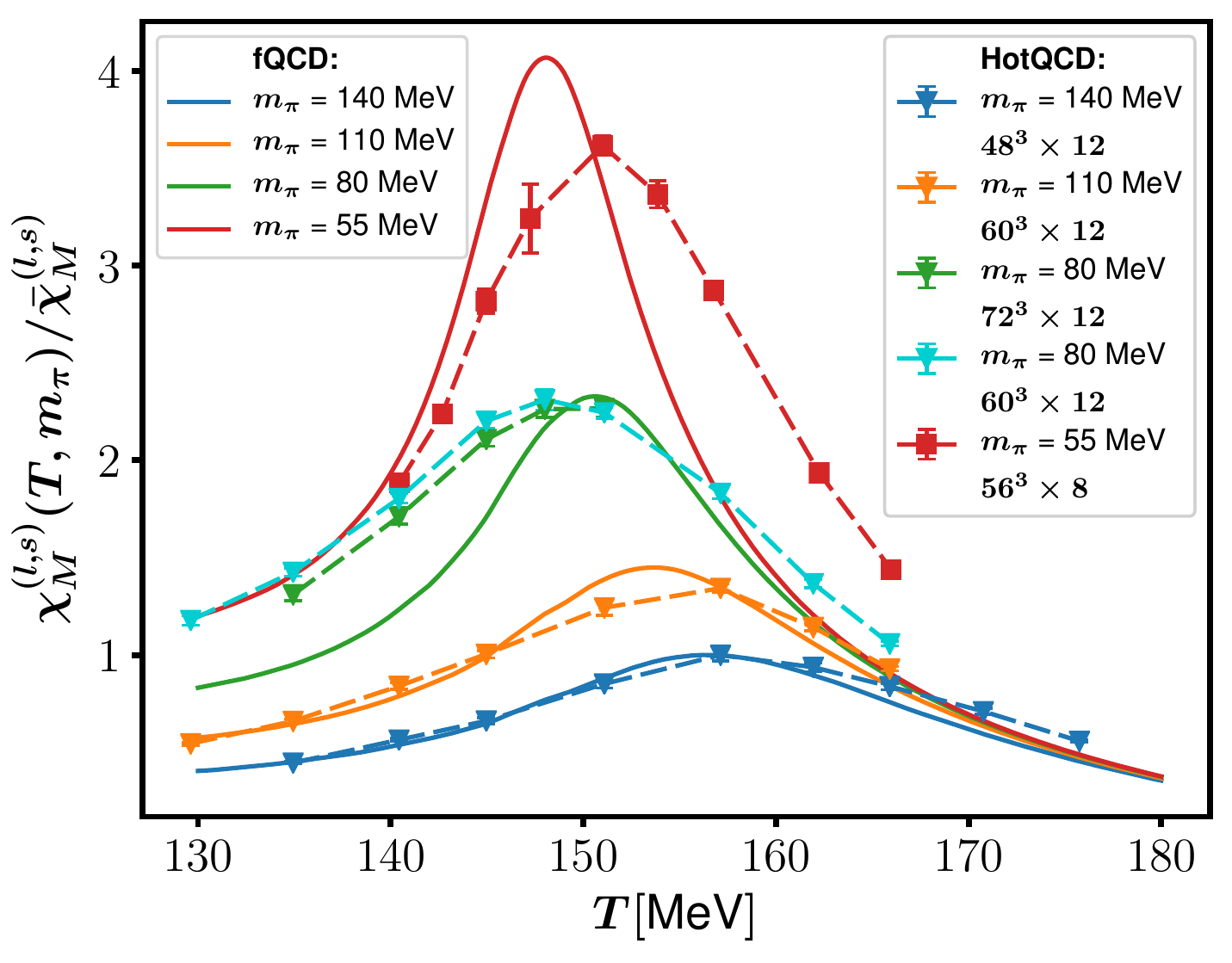}  
\caption{Left panel: Comparison of our fRG results for the pseudocritical temperature as a function of the pion mass 
to those from the HotQCD collaboration~\cite{Ding:2019prx}. The various dashed lines represent fits to the numerical data, see main text for details.
The estimates for the critical temperature~$T_{\text{c}}$ have been obtained from 
an extrapolation of the fits to~$m_{\pi}\to 0$. The temperatures $T_{60}^{(l,s)}$ and~$T_c^{\text{lattice}}$ are the extrapolated results for the chiral critical temperature 
obtained from a definition of the pseudocritical temperature which does not involve the peak position of the susceptibility, see main text for details. 
Right panel: Susceptibility as obtained from the reduced condensate as a function of the temperature. The  normalisation~$\bar\chi_M^{(l,s)}$ 
is the maximum of the susceptibility at the physical pion mass, see Eq.~\eq{eq:SusNorm}. The lattice QCD data 
has been taken from Refs.~\cite{Ding:2019prx,Ding:2019fzc}.}
\label{fig:tpcmpi}
\end{figure*}

The deviation of our estimate for the exponent~$p$ from the value  
associated with the $3d$ $O(4)$ universality class~\cite{Guida:1998bx,Kanaya:1994qe,Hasenbusch:2000ph} already 
suggests that QCD is not within the scaling regime, not even for the smallest pion masses considered 
in the present work. 
As already mentioned in Sect.~\ref{sec:intro}, this is in line with studies of low-energy effective theories of QCD. 
There, it has been found that deviations from scaling 
are still sizeable, even if the pseudocritical temperature scales approximately linearly for large pion masses
and also the suitably rescaled chiral susceptibilities for pion masses~$m_{\pi}\gtrsim 75\,\text{MeV}$ 
appear to fall approximately on one line~\cite{Braun:2010vd}. Even worse from a practical standpoint, 
actual scaling behavior of the chiral susceptibility and the chiral order parameter 
has only been observed for very small pion masses,~$m_{\pi}\lesssim 1\,\text{MeV}$~\cite{Braun:2010vd}. Given these results from low-energy QCD  
model studies and the fact that the exponent~$p$ is close to one for the $3d$ $O(4)$ universality class anyhow, 
a linear fit for the pseudocritical temperature may be considered reasonable. 
Performing such a linear fit, we obtain~$T_{\text{c}}\approx 142.4_{-0.1}^{+0.1}\,\text{MeV}$ 
from the extrapolation to~$m_{\pi}=0$, in very good agreement with our estimate for~$T_{\text{c}}$ presented above.

Let us now consider the ratio 
\begin{align}
D_{(l)}(m_{\pi})=\frac{T_{\text{pc}}^{(l)}(m_{\pi}) - T_{\text{c}}}{T_{\text{c}}}\,, 
\label{eq:DR}
\end{align}
which is an estimate for the relative dependence of the pseudocritical temperature on the pion mass. 
For the physical pion mass, $m_{\pi}=140\,\text{MeV}$, this ratio in our present 
first-principles fRG study is about a factor of three smaller than typical values for~$D_{(l)}$ found 
in low-energy QCD model studies~\cite{Berges:1997eu,Braun:2005fj}. For example, 
\begin{align}
D_{(l)}^{\text{QM}}(m_{\pi}\!=\!140\,\text{MeV})\approx 0.28
\end{align}
was reported in Ref.~\cite{Braun:2005fj} for the quark-meson (QM) model. 
In our present 
QCD study, we instead 
find
\begin{align}
D_{(l)}^{\text{QCD}}(m_{\pi}\!=\!140\,\text{MeV})\approx 0.10\,,
\end{align}
where we have employed the value for~$T_{\text{c}}$ 
obtained from an extrapolation of the pseudocritical temperature~$T_{\text{pc}}^{(l)}$ to the limit~$m_{\pi}=0$.

Next, we turn to the reduced susceptibility~$\chi_M^{(l,s)}$ as defined in Eq.~\eqref{eq:MagSus}. In Fig.~\ref{fig:renorm_cs} (right panel), we show a  
comparison of the light-quark susceptibility and the reduced susceptibility for three pion masses. 
As expected, the qualitative behaviour of the reduced susceptibility is the same as the one found for the light-quark susceptibility. More specifically, the 
susceptibilities increase for decreasing pion mass, indicating the approach to a singularity in the chiral limit. 
Fitting the relation~\eqref{eq:TpcRfct}  
to our numerical results for~$T_{\text{pc}}^{(l,s)}(m_{\pi})$ for~$m_{\pi}=30, 35, 40, \dots, 140\,\text{MeV}$, 
we obtain $T_{\text{c}} \approx 141.6_{-0.3}^{+0.3}\,\text{MeV}$, $c_{(l,s)} \approx 0.17_{-0.03}^{+0.03} \,\text{MeV}^{1-p}$,  
and~$p \approx 0.91_{-0.03}^{+0.03}$.  
Thus, the critical temperature~$T_{\text{c}}$ is in excellent agreement with the one extracted from our analysis of the light-quark susceptibilities, 
as it should be. With respect to the exponent~$p$, we note 
that it also deviates clearly from the expected $O(4)$ value. However, we observe that it is 
consistent within fit errors with the value for~$p$ which we obtained from our analysis of the light-quark susceptibility. 
Overall, we therefore cautiously
conclude that QCD is not within the scaling regime for the range of pion masses considered here, providing us with~$m_{\pi}\approx 30\,\text{MeV}$ 
as a conservative estimate for the upper bound of this regime. An actual determination of the 
size of the scaling regime is beyond the scope of present work as it requires to study very small pion masses. 
\renewcommand{\arraystretch}{1.5}
\begin{table*}
  \centering
 \begin{tabular}{|c|c|C{1.25cm}|C{1.25cm}|C{1.25cm}|C{1.25cm}|C{1.25cm}|C{1.25cm}|C{1.25cm}|C{1.25cm}|C{1.25cm}|C{1.25cm}|C{1.25cm}|}
    \hline
    \multicolumn{2}{|>{}c|}{}&\multicolumn{9}{c|}{$m_{\pi}$ [MeV]}\\
    \arrayrulecolor{kugray5}
    \arrayrulecolor{black}
    \cline{3-11}
    \multicolumn{2}{|>{}c|}{} &  $30$ & $40$ & $55$ & $70$ & $80$ & $100$ & $110$ & $120$ &  $140$ \\
    \hline
        \multirow{3}{*}{$T_{\text{pc}}$ [MeV]} & fQCD (reduced)  & $145.3$ & $146.4$ & $148.0$ & $149.6$ & $150.5$ & $152.7$ & $153.6$  & $154.8$ & $156.3$ \\
    \cline{2-11}
                         &HotQCD $(N_{\tau} = 12)$~\cite{Ding:2019prx}  & -- & -- & -- & -- & $149.7^{+0.3}_{-0.3}$ & -- & $155.6^{+0.6}_{-0.6}$ & -- & $158.2^{+0.5}_{-0.5}$ \\
    \cline{2-11}
                         &HotQCD $(N_{\tau} = 8)$~\cite{Ding:2019prx}  & -- & -- & $150.9^{+0.4}_{-0.4}$ & -- & $153.9^{+0.3}_{-0.3}$ & -- & $157.9^{+0.3}_{-0.3}$ & -- & $161.0^{+0.1}_{-0.1}$ \\
    \hline
  \end{tabular}
  \caption{Selection of peak positions of the reduced susceptibility for various pion masses as obtained from  
  our present fRG computation and a recent lattice QCD study~\cite{Ding:2019prx}. } 
  \label{tab:tc}
\end{table*}

In analogy to the definition~\eqref{eq:DR}, we can 
also define the relative dependence~$D_{(l,s)}(m_{\pi})$ 
of the pseudocritical temperature on the pion mass in 
case of the reduced susceptibility.
For~$m_{\pi}=140\,\text{MeV}$, we then find that this quantity is only slightly smaller than
the corresponding quantity associated with the light-quark susceptibility.

In Fig.~\ref{fig:tpcmpi} (right panel), we finally compare our fRG results for the reduced susceptibility to 
very recent results from the HotQCD collaboration~\cite{Ding:2019prx}. 
We observe excellent agreement between the results from the two approaches for pion masses~$m_{\pi}\gtrsim 100\,\text{MeV}$. 
The deviations of the results from the two approaches 
for smaller pion masses may at least partially be attributed to cutoff artefacts in the lattice data. 
Note that cutoff effects are expected to shift the maxima to smaller temperatures. We refer to
Ref.~\cite{Ding:2020rtq} for a respective discussion.

It is also worthwhile to compare the peak positions of the reduced susceptibilities extracted from the 
lattice QCD data with those 
from our fRG study, see Tab.~\ref{tab:tc} and~Fig.~\ref{fig:tpcmpi}~(left panel). 
As discussed above, the peak position can be used to define a pseudocritical temperature. 
For the presently available pion masses on the lattice, 
we find that the results from the two 
approaches for this pseudocritical temperature are in very good agreement. 
Moreover, we observe that at least a naive linear extrapolation of the HotQCD results  
for the peak position yields~$T_{\text{c}}\approx 144.6^{+0.5}_{-0.5}\,\text{MeV}$ 
for $N_{\tau}=8$ and~$T_{\text{c}}\approx 138.0^{+2.3}_{-2.3}\,\text{MeV}$ for~$N_{\tau}=12$, 
which is consistent with our estimate for~$T_{\text{c}}$.
However, as argued in Ref.~\cite{Ding:2019prx}, 
the strong pion-mass dependence of the so defined pseudocritical temperature potentially complicates the 
chiral extrapolation of lattice QCD data. Therefore, an alternative definition of the pseudocritical temperature 
has been introduced in Ref.~\cite{Ding:2019prx}. In the following we shall refer to 
this pseudocritical temperature as~$T_{60}^{(l,s)}$. Its implicit definition reads~\cite{Ding:2019prx}
\be
\chi_M^{(l,s)}(T_{60}^{(l,s)},m_{\pi})=0.6\max_T \chi_M^{(l,s)}(T,m_{\pi})\,.
\ee
Here, it is tacitly assumed that~$T_{60}^{(l,s)}$ is determined at 
a temperature to the left of the maximum of the susceptibility, implying~$T_{60}^{(l,s)} < T_{\text{pc}}^{(l,s)}$. 
For~$m_{\pi}\to 0$,~$T_{60}^{(l,s)}$ then converges 
to~$T_{\rm c}$. Moreover, 
the so defined pseudocritical temperature is expected to   
exhibit only a mild dependence on the pion mass and should hence 
be close to the chiral phase transition temperature~$T_{\rm c}$ for the range of pion masses of interest in the present work. 
In Ref.~\cite{Ding:2019prx}, this definition of the pseudocritical temperature has been used 
to extrapolate to the chiral limit, resulting in~$T_{\text{c}}^{\text{lattice}}=132^{+3}_{-6}\,\text{MeV}$. 
Employing this definition of the pseudocritical temperature to analyse 
our fRG results for the reduced susceptibility, we indeed observe an extremely weak dependence 
of~$T_{60}^{(l,s)}$ on the pion mass. To be specific, we find that it increases by less than~$1\,\text{MeV}$ 
when the pion mass is increased from~$m_{\pi}=30\,\text{MeV}$ to~$m_{\pi}=140\,\text{MeV}$. An extrapolation to the 
chiral limit yields~$T_{\rm c}\approx 142.4\,\text{MeV}$ which agrees nicely
with our estimates for~$T_{\text{c}}$ presented above, as it should be. Thus, from our fRG study, we eventually conclude 
\be
T_{\text{c}}\approx 142\,\text{MeV}\,
\ee
for the chiral phase transition temperature.  

\section{Conclusions}\label{sec:conc}
In this work, we have studied the magnetic susceptibility in $(2+1)$-flavour QCD within a first-principles fRG calculation. 
Specifically, we have presented results for the susceptibilities associated with the light-quark condensate 
and the reduced condensate. 

The chiral pseudocritical temperatures have been determined from the peak positions of the susceptibilities. 
Interestingly, we found that its dependence on the pion mass in the present QCD study is milder than in 
low-energy QCD model studies. From an extrapolation to the chiral limit, 
we obtained~$T_{\text{c}}\approx 142\,\text{MeV}$ for the chiral phase transition temperature. 

Our results for the susceptibilities and the scaling of the corresponding pseudocritical temperature 
indicate that QCD is not within 
the scaling regime for the considered pion masses $m_\pi\geq 30$\,MeV. As discussed in detail, this conclusion 
is at least in accordance with low-energy QCD model studies~\cite{Braun:2010vd}. There, a qualitatively similar behaviour of the susceptibilities 
and the pseudocritical temperature has been observed for the same pion mass range as considered here. However, the 
actual size of the scaling regime turned out to be significantly smaller. A detailed analysis of this issue within our present first-principles 
fRG approach is deferred to future work as it requires 
to study (very) small pion masses. 

We have also compared our results for the reduced susceptibility with very recent results from 
the HotQCD collaboration~\cite{Ding:2019prx} and found that the results 
from both approaches are in very good agreement for pion masses~$m_{\pi}\gtrsim 100\,\text{MeV}$.  For smaller pion masses, the 
lattice and fRG results are still consistent
with each other. Following the analysis of the HotQCD collaboration, 
we have also estimated the phase transition temperature in the chiral limit based on scaling properties of the susceptibility in the temperature 
regime below the temperature defined by the peak of the susceptibility. As it should be, this provides us with the same value for the chiral phase transition temperature 
as in the case of an extrapolation of the peak positions of the light-quark and reduced susceptibility. 

As a next step, it will be important to further extend our present study. For example, we plan to improve the 
stability of our numerical calculations for very small pion masses to eventually reach the chiral limit by employing
recent developments for the solution of fRG equations~\cite{Grossi:2019urj}.
Moreover, an analysis of the effect of the breaking of the~$U_{\rm A}(1)$ symmetry and effective~$U_{\rm A}(1)$ restoration at high temperatures is in order. 
A detailed analysis of the latter issue requires the study of Fierz-complete truncations of the effective action. 
First steps into this direction have been taken~\cite{BLPR2020}, suggesting that effective~$U_{\rm A}(1)$ restoration already sets in closely above 
the chiral phase transition temperature, in accordance with recent lattice QCD studies~\cite{Brandt:2019ksy}. An understanding  
of the effect of this almost coincidence of chiral and $U_{\rm A}(1)$ restoration 
on the critical behaviour is indeed an intriguing and not yet fully resolved problem in QCD.\\[-2ex]

\acknowledgments
We would like to thank the authors of Ref.~\cite{Ding:2019prx} for providing us with their lattice QCD data for 
the susceptibilities and for discussions. Moreover, the authors would like to thank O.~Kaczmarek and 
F.~Karsch for discussions and comments 
on the manuscript. 
JB acknowledges support by the DFG under Grant No. BR~4005/4-1 (Heisenberg program)
and by the Helmholtz International Center for the Facility for Antiproton and Ion Research (HIC for FAIR) within the LOEWE program of the State of Hesse.
JB and DR acknowledge support by the Deutsche Forschungsgemeinschaft (DFG, German Research Foundation) – Project number 315477589 – TRR 211. 
WF and SY are supported by
the National Natural Science Foundation of China under Contract No.~11775041. 
JMP is supported by EMMI, the BMBF grant 05P18VHFCA, 
and by the DFG Collaborative
Research Centre SFB 1225 (ISOQUANT) as well as by the DFG under
Germany's Excellence Strategy EXC - 2181/1 - 390900948 (the
Heidelberg Excellence Cluster STRUCTURES).  
FR is supported by the DFG through grant~RE~4174/1-1. 
This work is done within the fQCD collaboration~\cite{fQCD}.

%
\bibliography{qcd}

\end{document}